\documentclass[a4paper,aps,onecolumn,nofootinbib,preprintnumbers,superscriptaddress,amsmath,amssymb,amsfonts]{revtex4}
\usepackage{graphicx}
\usepackage{bm,natbib}
\usepackage{amsmath}
\usepackage[english]{babel}
\usepackage[T1]{fontenc}
\usepackage{amssymb}
\usepackage{amsfonts}
\usepackage{epsfig}
\usepackage{colordvi}
\usepackage{psfrag}
\usepackage{color}
\usepackage{ mathrsfs }
\newcommand{\Lagr}{\mathcal{L}}

\usepackage{dcolumn}
\usepackage{multirow}
\usepackage{hyperref}
\usepackage{epstopdf}
\usepackage{bm}
\usepackage{rotating}
\usepackage{lscape}
\usepackage{times}
\usepackage{comment}

\hypersetup{
  colorlinks=true,        
  linkcolor=blue,         
  citecolor=magenta,      
}

\begin{document}
\title{Minisuperspace quantum cosmology in $f(Q)$ gravity}

\author{Francesco Bajardi}
\email{f.bajardi@ssmeridionale.it}
\affiliation{Scuola Superiore Meridionale, Largo San Marcellino 10, I-80138, Naples, Italy.}
\affiliation{Istituto Nazionale di Fisica Nucleare (INFN) Sez. di Napoli, Compl. Univ. di Monte S. Angelo, Edificio G, Via Cinthia, I-80126, Napoli, Italy.}

\author{Salvatore Capozziello}
\email{capozziello@na.infn.it}
\affiliation{Scuola Superiore Meridionale, Largo San Marcellino 10, I-80138, Naples, Italy.}
\affiliation{Istituto Nazionale di Fisica Nucleare (INFN) Sez. di Napoli, Compl. Univ. di Monte S. Angelo, Edificio G, Via Cinthia, I-80126, Napoli, Italy.}
\affiliation{Dipartimento di Fisica ``E. Pancini'', Universit\`a di Napoli ``Federico II'',  Compl. Univ. di Monte S. Angelo, Edificio G, Via Cinthia, I-80126, Napoli, Italy.}

\date{\today}

\begin{abstract}
 $f(Q)$ symmetric-teleparallel gravity is considered in view of Quantum Cosmology. Specifically, we derive cosmological equations for $f(Q)$ models and then investigate the related energy conditions. In the Minisuperspace formalism, the point-like $f(Q)$ Hamiltonian is taken into account. In this framework, we obtain and solve the Wheeler-De Witt equation, thus finding the Wave Function of the Universe in different cases. We show that the Hartle Criterion can be applied  and classical observable universes occur.
\end{abstract}

\maketitle



\section{Introduction}
\label{introd}
Recent observations on supermassive black holes \cite{EventHorizonTelescope:2019dse, EventHorizonTelescope:2022xnr} and gravitational waves \cite{LIGOScientific:2016aoc} further confirm the validity of General Relativity (GR) as a theory describing gravitational phenomena also in the strong field regime. These results are complementary  to the well-known Solar System tests in the weak field limit. Despite these undoubted successes, Einstein's theory cannot be considered as the final theory for the gravitational interaction because it presents shortcomings at UV and IR scales \cite{Capozziello:2011et, Odintsov, Nojiri:2017ncd}. Furthermore, the impossibility to deal with GR under the same standard as the other interactions is one of the main weakness of the theory. Indeed, at  quantum level, UV divergences cannot be canceled out through standard renormalization techniques \cite{Bajardi:2021lwp, Goroff:1985th}. This problem has been firstly addressed by considering the asymptotic safety scenario \cite{Niedermaier:2006wt, Bonanno, Alessia}, according to which there should be nontrivial Gaussian fixed points in the renormalization group flow driving the values of the coupling constants in the UV regime. Another solution was proposed in Ref. \cite{Stelle:1976gc}, where the author extends the action including second-order curvature invariants and makes the resulting theory renormalizable, albeit the price to pay is the loss of unitarity. 

Although fixing UV problems in GR is one of the most active research areas nowadays, there is no final answer to the open questions occurring in GR, meaning that no self-consistent theory of Quantum Gravity so far exists. At the astrophysical and cosmological levels, incompatibilities with observations yielded the introduction of dark energy \cite{Copeland:2006wr} and dark matter \cite{Arkani-Hamed:2008hhe, Navarro:1995iw}, which, according to the current picture of the universe, should account for the majority of energy-matter content. The main puzzle is that, at the moment, any experimental attempt to address observed large-scale dark components as new fundamental particles has given no final result. 

The aforementioned problems are only few examples of the unsolved puzzles exhibited by GR \cite{Clifton:2011jh}. However, such difficulties led to the introduction of several different theories of gravity, modifying \emph{e.g.} the main assumptions of GR such as the Lorentz Invariance \cite{Horava:2009uw,  Carroll:2001ws, Caruana} or the Equivalence Principle \cite{Linder:2010py, Bengochea:2008gz, Cai:2015emx}, or extending the gravitational action to functions of higher-order curvature invariants \cite{Curve, Bajardi:2020osh, Nojiri:2005jg, Wheeler:1985nh, Bajardi:2022tzn, Capozziello:2021goa}. To the latter category belongs  the so called $f(R)$ gravity, which extends the Hilbert-Einstein action by including non-linear functions of the scalar curvature. This theory provides several interesting results at any scales. For instance, at galactic scales, a power-law model seems to be capable of fitting the galaxy rotation curves without any dark matter \cite{Capozziello:2006ph, Salucci, Borka, Boehmer:2007kx}. At cosmological scales, also dark energy can be mimicked without introducing any cosmological constant \cite{Starobinsky, Kamenshchik:2001cp, Scherrer:2004au, DAgostino:2021vvv, Benetti:2019lxu, Capozziello:2003gx}. Moreover, some models are potentially capable of addressing the mass-radius relation of Neutron Stars without introducing exotic equations of state (EoS) \cite{Farinelli, DeLaurentis:2018odx}, or the early stage of the inflationary universe without additional scalar fields \cite{Nojiri:2017ncd, Starobinsky:1980te}. At the quantum level, several models have been constructed with the purpose of addressing UV shortcomings of GR \cite{Becker:2006dvp, Ashtekar:2011ni, Modesto:2013jea, Witten:1995em, Capozziello:2021krv, Bajardi:2021hya}. 

Within the large amount of alternatives to Einstein's theory, particular interest has been recently gained by those theories involving torsion or non-metricity. Specifically, it can be shown that gravity can be described using torsion or non-metricity instead of curvature, and the resulting theories are dynamically equivalent to GR \cite{BeltranJimenez:2019esp, BeltranJimenez:2017tkd, Aldrovandi:2013wha, Maluf:2013gaa, Arcos:2004tzt, Ferrara}. As  pointed out in Sec. \ref{STEGRSec}, this is due to the fact that theories including either torsion scalar $T$ or non-metricity scalar  $Q$, instead of curvature scalar  $R$, yield exactly the same field equations as GR. As torsion is strictly related to the anti-symmetry of the affine connection with respect to the lowest indexes, non-metricity arise when considering a non-vanishing covariant derivative of  metric tensor, namely $\nabla_\mu g_{\alpha \beta} \neq 0$ where isometry is not conserved. The advantage of these approaches are, among the others, that Lorentz Invariance and Equivalence Principle are not required a priori like in GR. Furthermore, they can be dealt with within the framework of gauge theories \cite{Cai:2015emx}. These features could allow, if any, the possibility of a quantum approach to gravity avoiding shortcomings of GR. Furthermore, they seems to solve naturally several problems related to the cosmological dark side \cite{Saridakis, Finch:2018gkh, Mandal:2020lyq, Mandal:2020buf}.

In this paper we focus on the cosmological  extension of symmetric teleparallel  theory  whose action contains a function of the non-metricty scalar $Q$, \emph{i.e.} $f(Q)$. More precisely, we consider the function $f(Q) = \beta Q + \alpha Q^s$, with $\alpha, \beta$ and $s$ being free parameters. In this way, GR is straightforwardly recovered as soon as $\alpha \to 0$.

In the first part of the paper,  we consider a  Friedman-Lema\^itre-Robertson-Walker (FLRW) space-time with non-vanishing spatial curvature and find cosmological solutions of the field equations. Then we study the corresponding energy conditions and finally, recasting the model in the Hamiltonian formalism, we derive the Wheeler-De Witt (WDW) equation to obtain the so called \emph{Wave Function of the Universe}. Specifically, the latter approach is based on the  \emph{Arnowitt-Deser-Misner} (ADM) formalism \cite{DeWitt:1967yk, DeWitt:1967ub, Wheeler:1957mu, Bousso:2011up}, where the Hamiltonian  of GR is derived by means of a $(3+1)$ decomposition of the metric. The final result is a Schr\"odinger-like equation, the WDW equation, whose solution provides the Wave Function of the Universe. The meaning of the latter is still debated because  its probabilistic meaning,  as in the standard Quantum Mechanics, cannot be applied. However, several interpretations have been provided  \cite{Bousso:2011up, Vilenkin:1988yd, Hawking:1983hj, Vilenkin:1982de} and, despite the lack  of a direct  probabilistic interpretation,  it  gives important information on the  initial conditions and  evolution of the universe at early stages. Particularly relevant is the Hartle criterion, according to which an oscillating wave function can be related to classical dynamical systems interpretable as observable universes.

The paper is organized as follows:  Sec. \ref{STEGRSec} is a summary of  \emph{Symmetric Teleparallel Equivalent of General Relativity} (STEGR) and its modifications. In Sec. \ref{cosmostegrSec}, we consider the cosmology of the $f(Q) = \beta Q + \alpha Q^s$ model,  derive and solve  the corresponding field equations and study the related energy conditions.  The ADM formalism, introduced in Appendix \ref{ADMApp}, is then applied to the modified STEGR model in Sec. \ref{ADMSec}. The related WDW equation is solved in different cases. In particular, we show that the Hartle selection criterion (see Appendix  \ref{HartleSec}) holds and classical observable universes can be recovered. Conclusions are reported in   Sec. \ref{concl}.

\section{The Symmetric Teleparallel Equivalents of General Relativity}
\label{STEGRSec}
As mentioned in the introduction,  several possible extensions/modifications of GR can be proposed; a large  part of them relaxes some assumptions of GR. This is the case  of theories violating the Lorentz Invariance, or assuming general affine   connections which can be  non-metric compatible. See \cite{Ferrara} for a discussion. In the latter case, one of the possible modifications consists in  relaxing the assumption of symmetric connection with respect to the low indexes. As a consequence, the Equivalence Principle is not required at the foundation of the theory and torsion occurs in the description of the space-time dynamics \cite{Aldrovandi:2013wha}. In this way, the affine connection $\Gamma^\alpha_{\,\,\mu \nu}$ can be written in terms of the standard Levi-Civita connection $\hat{\Gamma}^\alpha_{\,\,\mu \nu}$ plus an additional contribution related to the torsion; clearly, GR is recovered as soon as the connection is assumed to be of the Levi-Civita one. In a space-time with non-vanishing curvature and torsion, the general connection can be written as
\begin{equation}
\Gamma^\alpha_{\,\,\mu \nu} = \hat{\Gamma}^\alpha_{\,\,\mu \nu} + K^\alpha_{\,\,\, \mu \nu} ,
\label{generalgamma}
\end{equation}
where $K^\alpha_{\,\,\,\mu \nu}$ is the \emph{contorsion tensor}  defined as
\begin{equation}
K^{\rho}_{\,\,\,\,\mu\nu}\equiv\frac{1}{2}g^{\rho\lambda}\bigl(T_{\mu\lambda\nu}+T_{\nu\lambda\mu}+T_{\lambda\mu\nu}\bigr), \quad \text{with} \quad T^\alpha_{\,\, \mu \nu} = 2 \Gamma^\alpha_{\,\,\,[\mu \nu]}. \label{contorsiontens}
\end{equation}
If one chooses the  \emph{Teleparallel Gauge}, where curvature vanishes identically, the gravitational action turns out to be equivalent to the Hilbert-Einstein one and the resulting theory is called \emph{Teleparallel Equivalent of General Relativity} (TEGR). In particular, setting $\Gamma^\alpha_{\,\,\mu \nu}  = 0$, Eq. \eqref{generalgamma} becomes
\begin{equation}
    \hat{\Gamma}^\alpha_{\,\,\mu \nu} = - K^\alpha_{\,\,\, \mu \nu} .
\end{equation}
Thus, by defining the \emph{torsion superpotential} and the \emph{torsion scalar}, respecively as
\begin{eqnarray}
S\,^{\rho \mu \nu} \equiv K\,^{\mu \nu \rho}-g^{\rho \nu} T\,_{\,\,\,\,\,\,\, \sigma}^{\sigma \mu}+g^{\rho \mu} T\,_{\,\,\,\,\,\,\, \sigma}^{\sigma \nu}\,, \label{superpot}
\end{eqnarray}
\begin{eqnarray}
&& T \equiv T\,^{\rho \mu \nu} S\,_{\rho \mu \nu}  \,, \label{definizione di torsione}
\end{eqnarray}
it turns out that the action
\begin{eqnarray}
S_{TEGR}\equiv -\frac{1}{2} \int d^4x\,\sqrt{-g}\, T \,,\label{TGACT}
\end{eqnarray}
is dynamically equivalent to the Hilbert-Einstein one and, therefore, it leads to the same field equations. It is worth noticing that, using tetrad fields, TEGR can be  recast as a gauge theory of  translation group in the flat tangent space-time \cite{Pereira:2019woq}

Similar considerations apply  relaxing another assumption of GR, that is the metricity principle. Imposing the covariant derivative of  metric to be non-vanishing, \emph{i.e.} $\nabla_\alpha g_{\mu \nu} \equiv Q_{\alpha \mu \nu}$, the affine  connection turns out to contain an additional contribution, called \emph{disformation tensor} and denoted by $L^{\alpha}_{\,\, \mu \nu}$. In a general space-time with curvature and non-metricity, the connection reads as
\begin{equation}
\Gamma^\alpha_{\,\, \, \mu \nu} = \hat{\Gamma}^\alpha_{\,\, \, \mu \nu} + \frac{1}{2}g^{\alpha\lambda}\bigl(-Q_{\mu \nu \lambda}-Q_{\nu \mu \lambda} + Q_{\lambda\mu\nu}\bigr) \equiv \hat{\Gamma}^\alpha_{\,\, \, \mu \nu} + L^\alpha_{\,\,\, \mu \nu}.
\end{equation} 
From the above definitions, it is possible to introduce the so called \emph{non-metricity scalar} $Q$ as
\begin{equation}
    Q \equiv -\frac{1}{4} Q_{\alpha \beta \gamma} Q^{\alpha \beta \gamma}+\frac{1}{2} Q_{\alpha \beta \gamma} Q^{\beta \gamma \alpha}+\frac{1}{4} Q_{\mu \,\,\,\, \lambda}^{\,\,\,\lambda} Q^{\mu \lambda}_{\quad \lambda}-\frac{1}{2} Q_{\mu \,\,\,\, \lambda}^{\,\,\,\lambda} Q_{\alpha}^{\, \, \, \, \mu \alpha},
\end{equation}
or, in a more compact form, as
\begin{eqnarray}
&& Q \equiv - \frac{1}{4} Q_{\alpha \mu \nu} \left[- 2 L^{\alpha \mu \nu}+  g^{\mu \nu} \left(Q^\alpha - \tilde{Q}^\alpha \right) - \frac{1}{2} \left(g^{\alpha \mu} Q^\nu + g^{\alpha \nu} Q^\mu  \right)\right], \label{Q1}
\end{eqnarray}
with
\begin{eqnarray}
&& Q_\mu \equiv Q_{\mu \,\,\,\, \lambda}^{\,\,\,\lambda}\,, \label{Q2}
\\ 
&& \tilde{Q}_{\mu} \equiv Q_{\alpha \mu}^{\, \, \, \, \, \, \alpha}\,. \label{Q3} 
\end{eqnarray}
Now,  to obtain a theory  dynamically equivalent to GR, one can choose the so called \emph{coincident gauge}, where the total affine connection vanishes, i.e.  $\Gamma^\alpha_{\,\, \, \mu \nu} = 0$. Therefore, the disformation tensor turns out to be equivalent to the Levi-Civita connection, up to a sign: 
\begin{equation}
L^\alpha_{\,\,\, \mu \nu} = - \hat{\Gamma}^\alpha_{\,\, \, \mu \nu} .
\label{coincgauge}
\end{equation}
As a consequence, the STEGR action, differing from the Hilbert-Einstein one by a boundary term, is
\begin{eqnarray}
&&S_{STEGR}\equiv \frac{1}{2} \int d^4x\,\sqrt{-g}\,Q\,. \label{stegraction}
\end{eqnarray}
This is due to the fact that the scalar curvature (written in terms of the Levi-Civita connection) and the non-metricity scalar are equivalent up to a total divergence. Specifically, considering Eqs. \eqref{Q1}-\eqref{Q3}, one gets
\begin{equation}
\hat{R}= -Q - \hat{\nabla}_{\alpha}\left(Q^{\alpha} + \tilde{Q}^{\alpha}\right),
\label{Q + B}
\end{equation}
with $\hat{\nabla}_{\alpha}$ being the covariant derivative written in terms of the Levi-Civita connection.

Also notice that, in the coincident gauge, the non-metricity scalar can be  written in terms of the Levi-Civita connection and the metric as \cite{BeltranJimenez:2019esp}:
\begin{equation}
    Q_{CG} = g^{\mu \nu}\left(\hat{\Gamma}^{\alpha}{ }_{\beta \mu} \hat{\Gamma}^{\beta}{ }_{\nu \alpha}-\hat{\Gamma}^{\alpha}{ }_{\beta \alpha} \hat{\Gamma}^{\beta}{ }_{\mu \nu}\right).
\end{equation}
According to  the above considerations, the gravitational interaction can be equivalently described either by curvature, or by torsion, or by non-metricity. 

In summary, the general connection containing all possible contributions reads  
\begin{equation}\Gamma^{\alpha}_{\,\, \mu \nu} = \hat{\Gamma}^{\alpha}_{\,\, \mu \nu} + K^{\alpha}_{\,\, \mu \nu} + L^{\alpha}_{\,\, \mu \nu}\,.
\end{equation}
 Einstein's gravity is recovered for  $K^{\alpha}_{\,\, \mu \nu} + L^{\alpha}_{\,\, \mu \nu} = 0$, so that  connection is  Levi-Civita. In the {\it teleparallel gauge}, the Levi-Civita contribution turns out to be equivalent to the contorsion tensor, so that the space-time can be described by torsion only. In the {\it coincident gauge}, where $K^{\alpha}_{\,\, \mu \nu} = \Gamma^{\alpha}_{\,\, \mu \nu} =0$, non-metricity is the only non-vanishing component. The actions corresponding to these three theories are totally equivalent up to a boundary term. This  means  that the related field equations are exactly the same. 

\subsection{The $f(Q)$ extension}
\label{f(Q)ext}
As pointed out in \cite{Heisenberg:2018vsk}, non-metricity, torsion and curvature can be thought as a geometric \emph{Trinity of Gravity}, since all of them leads to totally equivalent theories by different formalism. Nonetheless, the equivalence among the three models also implies that TEGR and STEGR suffer the same shortcomings as GR at cosmological and astrophysical levels. For this reason, in analogy to GR extensions like $f(R)$ gravity, $f(T)$ and $f(Q)$ extensions are considered in the literature \cite{Wu:2010mn, Krssak:2015oua, BeltranJimenez:2019tme, Lazkoz:2019sjl, Bajardi:2020fxh, Capriolo1, Capriolo2, Ferrara, Calza:2022mwt}. 
However, it is worth pointing out that  actions involving $f(R)$, $f(T)$, and $f(Q)$
are not equivalent, though they can be made equivalent by introducing the corresponding boundary terms \cite{Caruana:2020szx, Bahamonde:2021srr, Bahamonde:2016grb}. As a matter of fact, $f(R)$ gravity is different with respect to the $f(T)$ and $f(Q)$ theories, since the corresponding boundary terms $B_T$ and $B_Q$ play a non-trivial role in the dynamics. 

For instance,  $f(T)$ and $f(Q)$ gravity lead to second-order field equations \cite{Cai:2015emx}, while $f(R)$ gravity gives fourth-order field equations. Moreover, under proper assumptions, $f(T)$ and $f(Q)$ models can avoid bad ghosts and lead to unitary theories \cite{Conroy:2017yln}. Interestingly, equivalences among the three theories can occur when considering the so called Noether Symmetry Approach \cite{Book, Cimento, Dialektopoulos:2018qoe, Urban:2020lfk}. Specifically, when the actions are selected by Noether symmetries, it turns out that the related symmetry generators can be made equivalent under proper definitions of the free parameters \cite{Bajardi:2020xfj}.

The features of $f(T)$ and $f(Q)$ gravity are today investigated in the literature through the applications to cosmology and astrophysics. 

For instance,  in \cite{Finch:2018gkh},  authors show that a $f(T)$ power-law model can fit the galaxy rotation curve. In \cite{Aljaf:2022fbk}, the $H_0$ tension is addressed. In \cite{Bamba:2010wb}, an EoS, derived from $f(T)$ can address the dark energy issue. In \cite{Capozziello:2017uam} a model-independent way to solve the modified Friedman equations, in the framework of $f(T)$ cosmology, is proposed. 

Regarding $f(Q)$ gravity, in Ref. \cite{Anagnostopoulos:2022gej} the Big Bang Nucleosynthesis is studied, while, in \cite{Shokri}, slow-roll inflation is considered. In \cite{Bajardi:2020fxh},   bouncing cosmological models are considered within the context of $f(Q)$ gravity. In \cite{Soudi:2018dhv},  authors investigate  the polarization of  gravitational waves. In \cite{Anagnostopoulos:2021ydo}, $f(Q)$ gravity is showed to be potentially capable of challenging the $\Lambda$CDM model and, in \cite{Rocco}, a model-independent reconstruction of $f(Q)$ cosmology is reported. In \cite{Banerjee:2021mqk}, wormhole solutions are provided by static and spherically symmetric backgrounds. To conclude a wide class of phenomena and models can be investigated  under the standard  of $f(Q)$ formalism.

Considering dynamics generated only by non-metricity, the variation of  $f(Q)$ action yields the field equations \cite{Dialektopoulos:2019mtr}
\begin{equation}
\begin{split}
& \frac{2}{\sqrt{-g}} \hat{\nabla}_{\alpha}\left(\sqrt{-g} g_{\beta \nu} f_{Q} \left[ L^{\alpha \mu \beta}- \frac{1}{2} g^{\mu \beta} \left(Q^\alpha -  \tilde{Q}^\alpha \right) + \frac{1}{4} \left(g^{\alpha \mu} Q^\beta + g^{\alpha \beta} Q^\mu  \right)\right]\right)  \\
&+ f_{Q} \left[ L^{\mu \alpha \beta}- \frac{1}{2} g^{\alpha \beta} \left(Q^\mu -  \tilde{Q}^\mu \right) + \frac{1}{4} \left(g^{\mu \alpha} Q^\beta + g^{\mu \beta} Q^\alpha  \right)\right] Q_{\nu \alpha \beta} - \delta_{\nu}^{\mu} f =\mathcal{T}_{\,\,\,\nu}^{\mu}\,, \\ &
\end{split}
\label{FE f(Q)}
\end{equation} 
with $\mathcal{T}_{\,\,\,\nu}^{\mu}$ being the matter stress-energy tensor. 

Here we want to derive exact solutions of  $f(Q)$ field equations, in a cosmological background with non-vanishing spatial curvature. After, we will develop the Quantum Cosmology formalism to find out the Wave Function of the Universe. Such a wave function, according to the Hartle criterion, allows to recover the classical evolution of observable universes.

\section{Cosmology in $f(Q)$ Symmetric Teleparallel Gravity}
\label{cosmostegrSec}
Let us consider the modified symmetric teleparallel action in a cosmological background. We are going to adopt the Lagrange Multipliers method to find a point-like Lagrangian and the related equations of motion. 
The Lagrange multipliers method is a general mathematical technique used to deal with constraints into  optimization or variational problems. In the context of modified gravity, the Lagrange multipliers emerge as  constraints highlighting possible further degrees of freedom of theories different with respect to GR.
Specifically, by introducing them, one can construct  modified action functionals. Variations of these  actions with respect to the Lagrange multipliers and the canonical variables  yield the modified equations of motion. Lagrangians  can be determined through this procedure, as they are derived from the modified action functionals \cite{Capozziello:2010uv}.
It is worth to point out that the specific form of the Lagrangian, obtained by the Lagrange multipliers, strictly depends on the particular theory of  gravity. In the specific case of this paper, we are going to consider the cosmological expression of the non-metricity scalar, which infers the Lagrange multiplier $\lambda$, appearing as a constraint in the point-like Lagrangian.
To this purpose, let us start from the action
\begin{equation}
S = \int \sqrt{-g} f(Q) \, d^4 x,
\end{equation}
containing a function of the non-metricity scalar $Q$. We can get a point-like Lagrangian considering  a FLRW metric of the form
\begin{equation}
ds^2 = dt^2 - a(t)^2 \left[\frac{ dr^2}{1 - k r^2}  + r^2 d\Omega_{2}^2\right],
\label{intervalstarting}
\end{equation}
with $k$ being the spatial curvature and $a$ the scale factor. Using the Lagrange multipliers method and the cosmological expression of $Q$ with respect to the space-time \eqref{intervalstarting}, the action, up to a total factor,  can be recast as
\begin{equation}
S = 2\pi^2\int \left[a^3 f(Q) - \lambda \left(Q - 6\frac{k}{a^2} +6\frac{\dot{a}^2}{a^2}\right) \right]dt.
\end{equation}
The Lagrange multiplier $\lambda$, as standard, can be found by varying the action with respect to the non-metricity scalar, providing
\begin{equation}
    \frac{\delta S}{\delta Q}  = 0 \to \lambda = a^3 f_Q(Q),
\end{equation}
then the cosmological point-like Lagrangian turns out to be
\begin{equation}
\Lagr = \left[a^3 (f - Q f_Q) + 6 a (k - \dot{a}^2) f_Q  \right].
\end{equation}
From the above Lagrangian it is possible to obtain three partial differential equations, namely the two Euler-Lagrange equations with respect to $a$ and $Q$:
\begin{eqnarray}
       && \frac{d}{dt} \frac{\partial \Lagr}{\partial \dot{a}} - \frac{\partial \Lagr}{\partial a} = 0, \label{ELa1}
        \\
        && \frac{d}{dt} \frac{\partial \Lagr}{\partial \dot{Q}} - \frac{\partial \Lagr}{\partial Q} = 0 \label{ELQ1}.
\end{eqnarray} 
The system must be implemented by the so called energy condition, \emph{i.e.}
\begin{equation}
         E_{\Lagr}=\dot{a} \frac{\partial \Lagr}{\partial \dot{a}} + \dot{Q} \frac{\partial \Lagr}{\partial \dot{Q}} - \Lagr = 0\,. \label{EC1}
   \end{equation}
It is important to stress that dynamics gives the evolution of the scale factor. This is because the equation with respect to the non-metricity scalar is a constraint imposed by the Lagrange multiplier. Furthermore, the zero energy condition, directly inferred from the Einstein equations, is another constraint completing the dynamical system. In particular, the vanishing energy condition implies that the related Hamiltonian is zero on the constraint surface due to time reparameterization invariance \cite{Agrawal:2020xek}. As a result, the variational approach, naturally providing two different Friedman equations by the variation with respect to the metric, turns out to be completely equivalent to the Lagrangian approach.

On the other hand, the equation with respect to the scale factor corresponds to the (1,1) Einstein equation, while the Euler-Lagrange equation with respect to $Q$ provides  the cosmological expression of the non-metricity scalar. Specifically, being
\begin{equation}
 \frac{\partial \Lagr}{\partial \dot{Q}} =0\,,
 \end{equation}
 we have
 \begin{equation}
 \frac{\partial \Lagr}{\partial Q} = 0\, \quad \to \,\, \left[a^2 Q - 6 (k-\dot{a}^2) \right] f_{QQ} = 0\,
 \end{equation}
 and then 
\begin{equation}
    Q = 6 \left(\frac{k}{a^2} - \frac{\dot{a}^2}{a^2}\right)\,,
\end{equation}
which is the definition of the non-metricity scalar in FLRW space-times. The situation is completely equivalent to that in $f(T)$ teleparallel equivalent gravity, where the point-like Lagrangian results reduced and independent of the time derivative $\dot{T}$, with $T$ being the torsion scalar \cite{Basilakos:2013rua}. This reduction mechanism, related to the Lagrange multipliers, is present in several modified gravity models, see \emph{e.g.} \cite{Book}. 

Albeit, in principle, the non-metricity scalar can be written as a function of $a$ and $\dot{a}$, the Lagrangian formalism implies that $Q$ and $a$ have to be treated as separated fields and their relation is thus recovered by the Euler-Lagrange equation with respect to $Q$, namely Eq. \eqref{ELQ1}. This approach is standard in the literature (see \emph{e.g.} \cite{Bahamonde:2019swy, Bahamonde:2018zcq, Capozziello:2007wc, Sharif:2014fla}) and will be used to reduce dynamics.

Eqs. \eqref{ELa1} and \eqref{EC1} read, respectively, as
\begin{equation}
    \begin{split}
        &2 \left(\dot{a}^2+k\right) f_Q+ 4 a \left(\ddot{a} f_Q +\dot{a} \dot{Q} f_{QQ} \right)+a^2 \left(f-Q f_Q\right)=0\,,
   \\
   &6 \left(\dot{a}^2+k\right) f_Q+a^2 \left(f-Q f_Q\right) =0.
    \end{split}
    \label{FEfQ}
    \end{equation}
  For $k=0$, the above equations reduce to
  \begin{eqnarray}
&& 2 Q f_Q - f = 0\,,   
\\
&& 2 Q f_{QQ} + f_Q = 0\,.
\end{eqnarray}
Let us now consider the function $f(Q) = \beta Q + \alpha Q^s$, with $s, \alpha, \beta$, real constants. It is the straightforward generalization of STEGR. The Lagrangian becomes
\begin{equation}
   \Lagr=  \left[6 a \left(k-\dot{a}^2\right) \left(\alpha s Q^{s-1}+\beta\right)+\alpha  (1-s)
   a^3 Q^s\right].
   \label{specificlagra}
\end{equation}
Starting from Eq. \eqref{specificlagra}, we can  find cosmological solutions with generic spatial curvature $k$ and, therefore, study the energy conditions in terms of the free parameters.
\subsection{Cosmological Solutions}
For $\alpha$ and $\beta$ different from zero, the general solution of the Euler-Lagrange equations can be found analytically only for $k=0$. We have
\begin{equation}
    a(t) = a_0 e^{\ell t} \quad Q(t) = -6 \ell^2,
    \label{cosmosolgeneral}
\end{equation}
with 
\begin{equation}
 \ell =   \frac{1}{\sqrt{6}}\left(\frac{\alpha (2s-1)}{\beta }\right)^{\frac{1}{2-2 s}}.
\end{equation}
Considering $s = 2$, we have  a Starobinsky equivalent model in non-metric gravity with a scale factor of the form
\begin{equation}
\label{Staro}
    a(t) = e^{\sqrt{\frac{\beta}{18 \alpha}} \, t}  .
\end{equation}
The related inflationary model has been studied in \cite{Shokri}. In order to obtain an accelerating behavior, we must set $\beta/\alpha > 0$. Notice, moreover, that the solution \eqref{cosmosolgeneral} is obtained by imposing $s \neq 1, 1/2$. As a matter of fact, for the latter value, field equations are identically satisfied independently of the scale factor. The $\Lambda$CDM model is clearly recovered for $s = 0$ and $\beta = -1/2$, where Eq. \eqref{cosmosolgeneral} can be recast as
\begin{equation}
    a(t) = a_0 e^{\sqrt{\alpha/3} \,  t}.
\end{equation}
\subsubsection{The Case $\alpha = 0$}
Analytic solutions with non-vanishing spatial curvature are present only if either $\alpha = 0$ or $\beta = 0$. In the former case,  we have
\begin{equation}
    a(t) = c_1+\sqrt{-k} \, t \quad Q(t) = \frac{12 k}{\left(c_1+\sqrt{-k} t\right){}^2},
    \label{alpha0solutions}
\end{equation}
with $c_1$ being an integration constant. As in GR, when the spatial curvature vanishes, the cosmological scale factor becomes trivial in vacuum.
\subsubsection{The Case $\beta = 0$}
Let us now solve  system \eqref{FEfQ} by setting $\beta = 0$. The only analytic scale factor solving the field equations is
\begin{equation}
    a(t) = c_1+\frac{\sqrt{k} \, t}{\sqrt{2s-1}}\,, \qquad Q(t) = -\frac{12 k s}{\left(c_1 \sqrt{2s-1}+\sqrt{k} \, t\right){}^2}\,.
    \label{a(t)beta0}
\end{equation}
If $k = 0$, only trivial scale factors occur as solutions of the cosmological equations. Moreover, in the special case $s=1/2$, the Euler-Lagrange equations turn out to be trivially satisfied for any scale factor. Also notice that this solution only depends on the spatial curvature $k$, unlike $f(R) = \alpha R^s$ gravity, where the cosmological dynamics is determined by the value of the parameter $s$. In both cases, GR is recovered for $s = 1$ and little deviations can be studied by setting $s = 1+\epsilon$, with $\epsilon \ll 1$. In this case, the function $f(Q) = \alpha Q^{1+\epsilon}$ can be expanded around $\epsilon = 0$, providing
\begin{equation}
    Q^{1+\epsilon} \sim Q + \epsilon \, Q \ln Q
\end{equation}
and the additional logarithmic contribution can behave like an effective cosmological constant. 
\subsection{Energy Conditions}
The energy condition for this model can be studied for different values of the free parameters $s$ and $\alpha$. In order to recover the gravitational coupling, we set $\beta = -1/2$, so that the RHS of the field equations \eqref{FEfQ} can be though as effective energy density $\rho$ and pressure $p$ provided by geometry, whose expressions are
\begin{equation}
    \begin{split}
  &  \rho =   (1-2s) \alpha Q^s 
  \\
  & p = (2s-1) \alpha Q^{s-1} (Q + 4 s \dot{H}),
    \end{split}
    \label{rhopress}
\end{equation}
with $H = \dot{a}/a$ being the Hubble parameter. Therefore, the EoS parameter $w$ reads
\begin{equation}
  w = p/\rho =-1  -\frac{4 s \dot{H}}{Q}
\end{equation}
Notice that by plugging the solution \eqref{cosmosolgeneral} into Eq. \eqref{rhopress}, the energy density and the pressure turn out to be constant and opposite. Specifically, they are
\begin{equation}
  \rho = -p= \alpha  (1-2 s) \left[2\alpha  (2 s-1)\right]^{\frac{s}{1-s}}.
\end{equation}
Hence, the EoS parameter is constantly equal to $-1$, independently of the time in which it is evaluated. This means  that the additional contribution $\alpha Q^s$  behaves like an effective cosmological constant.  

Clearly, in this case, the Null Energy Condition and the Dominant Energy Condition vanish identically, while the Weak Energy Condition and the Strong Energy condition can be violated depending on the values assumed by $s$ and $\alpha$. For instance, choosing a coupling constant ranging from 0 up to 20 time the GR coupling constant, the WEC is always violated, meaning that $f(Q)$ gravity can act as an exotic fluid. In other  words, the further gravitational degrees of freedom resulting from $f(Q)$ affect dynamics giving a further fluid to source the field equations. See also \cite{Lobo1, Lobo2} for a discussion on generalized energy conditions

\section{$f(Q)$ quantum cosmology}
\label{ADMSec}
Quantum Cosmology is an important tool to investigate  features of the early universe. In particular, it is an approach needed to fix the initial conditions from where dynamical systems, describing observable universes, come out after some prescriptions are satisfied   \cite{Vilenkin:1988yd, Hawking:1983hj,Vilenkin:1982de}. In view of reproducing the standard of Quantum Mechanics results, the Hamiltonian formalism and a quantization procedure have to be pursued. 

The Quantum Cosmology approach can  be  applied to different theories of gravity \cite{Bajardi:2021tul, Capozziello:2022vyd, Capozziello:2012hm} and results are particularly useful in the framework of the so called \textit{Minisuperspace} formalism   \cite{QuantumIJGMMP}. Minisuperspaces are suitable reductions of the 3-metrics configuration space (the \textit{Superspace}) where cosmological dynamics can be constructed.

The application of this procedure   allows to obtain the WDW  equation from the related Hamiltonian and then the so called  \textit{Wave Function of the  Universe}, a quantity connected to the probability to achieve  configurations of cosmological variables  which can be interpreted (or not) as initial conditions of the universe (or of the "universes" in a Many Worlds Interpretation of Quantum Mechanics). 

Despite the meaning of the Wave Function of the Universe remains uncertain, numerous explanations have been proposed throughout the years. For instance, according to the Many World Interpretation, the wave function originates from quantum measurements that occur simultaneously in different universes, without any collapse of the wave function \cite{Bousso:2011up}. Another perspective was presented by Hawking, who suggested that the wave function is connected to the probability that early universe evolves into our classical universe \cite{Vilenkin:1988yd, Hawking:1983hj}. In \cite{Vilenkin:1982de, Vilenkin:1984wp}, the emergence of the universe from "nothing" is explored, viewing the universe from a quantum standpoint. This viewpoint helps in solving the challenge of determining the boundary condition, where it is asserted that the fundamental laws of physics must be intrinsic and not imposed externally \cite{Vilenkin:1986cy}. In this interpretative framework, J. B. Hartle proposed a criterion for extracting information from the wave function based on its behavior in the late-time epoch. Specifically, according to the Hartle criterion, the wave function should exhibit oscillatory behavior within the classically allowed region. Oscillations of the wave function mean that cosmological parameters are correlated. As stated by Hartle, this feature corresponds to describe classical universes \cite{Hartle:1983ai}. This interpretation of the wave function is analogue to that provided by non-relativistic Quantum Mechanics, where the solution of the Schr\"odinger equation for a particle inside a potential barrier, \emph{i.e.} in the classically forbidden region, yields a decaying exponential behavior while the wave function outside the barrier  gives rise to an oscillating behavior. Adopting the Hartle criterion, in the semiclassical limit, allows us to express the wave function in terms of the classical action $S$ as $\psi \sim e^{i S}$ and then deriving classical cosmological solutions (See the Appendix \ref{HartleSec} for details).

In the specific case of $f(Q)$ cosmology, after a Legendre transformation of Lagrangian \eqref{specificlagra}, the related Hamiltonian is
\begin{equation}
\mathcal{H} = -6 k a \left(\beta +\alpha  s Q^{s-1}\right)-\frac{\pi_a^2}{24 a \left(\beta
   +\alpha  s Q^{s-1}\right)}+\alpha  (s-1) a^3 Q^s,
   \label{Hamilt}
\end{equation}
where
\begin{equation}
\pi_a = \frac{d \Lagr}{d \dot{a}} = - 12 a \dot{a} f_Q\,,
\end{equation}
is the conjugate momentum related to the scale factor $a(t)$. Notice that there is no momentum related to the non-metricity scalar, due to the reason discussed in Sec.\ref{cosmostegrSec}. In fact, though the Minisuperspace $\mathcal{S} = \{a,Q\}$ is two-dimensional, the resulting tangent space $\mathcal{TS} = \{a,\dot{a},Q \}$ is three-dimensional. 

The Hamiltonian in Eq. \eqref{Hamilt} encompasses the two variables $a$ and $Q$, treated as separate quantities. Their correlation is reestablished through the Hamilton-Jacobi equations, equivalent to the Euler-Lagrange equations by construction. In this scenario, adopting the Lagrangian formalism becomes crucial, as the Hamiltonian can be simply derived through a straightforward Legendre transformation and the problem of infinite-dimensional Superspace can be overcome by a related Minisuperspace (see Appendix \ref{ADMApp} for details).

In the general case, that is for $\alpha, \beta \neq 0$, the WDW equation can be exactly solved only by setting $k = 0$. In such a case, the Hamiltonian turns out to be
\begin{equation}
    \mathcal{H} = \alpha  (s-1) a^3 Q^s-\frac{\pi_a^2}{24 a \left(\beta +\alpha  s Q^{s-1}\right)}\,,
\end{equation}
and by solving the differential equation $\mathcal{H} \psi = 0$, one gets
\begin{equation}
     \psi(a,Q) =  \psi_0 \, \sqrt{a} \, p(Q)^{1/6} \left[c_1 J_{\frac{1}{6}}\left(a^3 p(Q)\right) +c_2 J_{-\frac{1}{6}}\left(a^3 p(Q)\right)\right] ,
     \label{WFcomplete}
\end{equation}
where we have defined
\begin{equation}
  p(Q) \equiv \sqrt{\alpha  (1-s) Q^{s-1}\left(\alpha  s Q^s+\beta  Q\right)}\,.  
\end{equation}
Here $c_1$ and $c_2$ are dimensionless complex coefficients and $J_{\pm\frac{1}{6}}\left(a^3 p(Q)\right)$ are Bessell functions of first kind. Setting, for instance, $s=2$,  we have $f(Q) = \beta Q + \alpha Q^2$, and then
\begin{equation}
\label{Staro1}
    \psi(a,Q) =  \psi_0 \, \sqrt{a} \, \left[-\alpha Q^2 \left( 2\alpha  Q+\beta\right) \right]^{1/12} \left\{c_1 J_{\frac{1}{6}}\left[a^3 \left(\sqrt{-\alpha Q^2 \left( 2 \alpha  Q+\beta\right)}  \right)\right] +c_2 J_{-\frac{1}{6}}\left[a^3 \left(\sqrt{-\alpha Q^2 \left(2 \alpha  Q+\beta\right)}  \right)\right]\right\}\,. 
\end{equation}
This is the symmetric teleparallel analogue of the Starobinsky model whose inflationary counterpart has been studied in Ref.\cite{Shokri}.
It is  worth  noticing that, for large values of $a$, the wave function keeps an oscillating behaviour, while in the early stages its trend appears exponential-like.  The plot of the Wave Function of the Universe in terms of $a$ is reported in Fig. 1. This is in agreement with the Hartle Criterion \cite{Halliwell:1989dy} by which it is possible to recover the classical behavior of the universe. Specifically, the first step is to adopt the semiclassical limit \cite{Maniccia:2022iqa} and recast the wave function as $\psi \sim e^{i S_0}$, with $S_0$ being the Hamilton-Jacobi action. Since we are interested in finding classical solutions, we can take the limit of Eq. \eqref{WFcomplete} in the late times, where the combination of the Bessel functions is oscillating. Specifically, up to total factors, the classical action can be identified as the argument of the Bessel functions, namely 
\begin{equation}
S_0 = c_0 a^3 p(Q), 
\end{equation}
where $c_0$ is a constant parameter. Formally, from the above action, we get the conjugate momenta:
\begin{equation}
\pi_a = \frac{\partial S_0}{\partial a} = 3c_0 a^2 p(Q), \qquad \pi_Q = \frac{\partial S_0}{\partial Q} = \frac{\alpha  (1-s) Q^{s-1} \left(\beta +\alpha  s^2 Q^{s-1}\right)+\alpha  (1-s) (s-1)
   Q^{s-2} \left(\alpha  s Q^s+\beta  Q\right)}{2 \sqrt{\alpha  (1-s) Q^{s-1}
   \left(\alpha  s Q^s+\beta  Q\right)}}\,,
\end{equation}
where, from the discussion in Sec. \ref{cosmostegrSec}, it results $\pi_Q=0$.

Comparing the above momenta with those provided by Lagrangian \eqref{specificlagra}, we obtain the following  equations:
\begin{equation}
  3c_0 a^2 p(Q) = -12 a \dot{a} (\alpha s Q^{s-1} + \beta),
\end{equation}
\begin{equation}
\label{EJQ}
    \alpha  (1-s) Q^{s-1} \left(\beta +\alpha  s^2 Q^{s-1}\right)+\alpha  (1-s) (s-1)
   Q^{s-2} \left(\alpha  s Q^s+\beta  Q\right) = 0,
\end{equation}
where Eq. \eqref{EJQ} is an algebraic constraint resulting from the degenerate momentum $\pi_Q=0$.
The cosmological solution is
\begin{equation}
 a(t) = a_0  \exp \left(-\frac{c_0 t \sqrt{\alpha  (s-1) \left(-Q^{s-1}\right) \left(\alpha  s
   Q^s+\beta  Q\right)}}{4 (\beta +2 \alpha  Q)}\right), \qquad Q(t)= -\left(\frac{\alpha  \beta  s-\alpha  \beta  s^2}{-2 \alpha ^2 s^3+3 \alpha ^2 s^2-\alpha
   ^2 s}\right)^{\frac{1}{s-1}},
\end{equation}
in agreement with those obtained in Sec.\ref{cosmostegrSec} where the scale factor evolves exponentially and the non-metricity scalar is a constant. Moreover, from the relation occurring between the scale factor and the non-metricity scalar, it is possible to constrain the parameter $c_0$.  

It is worth noticing that, using the Hartle Criterion, we can recover classical solutions by asking for the existence of Noether symmetries, see \emph{e.g.} \cite{Bajardi:2021tul, Capozziello:2022vyd, Bajardi:2020osh, Gaetano}. In this case, such a criterion demonstrates to be a selection rule. See Appendix \ref{HartleSec}.

Let us now focus on the cases where either $\alpha = 0$ or $\beta = 0$.
\begin{center}
\centering
\includegraphics[width=.55\textwidth]{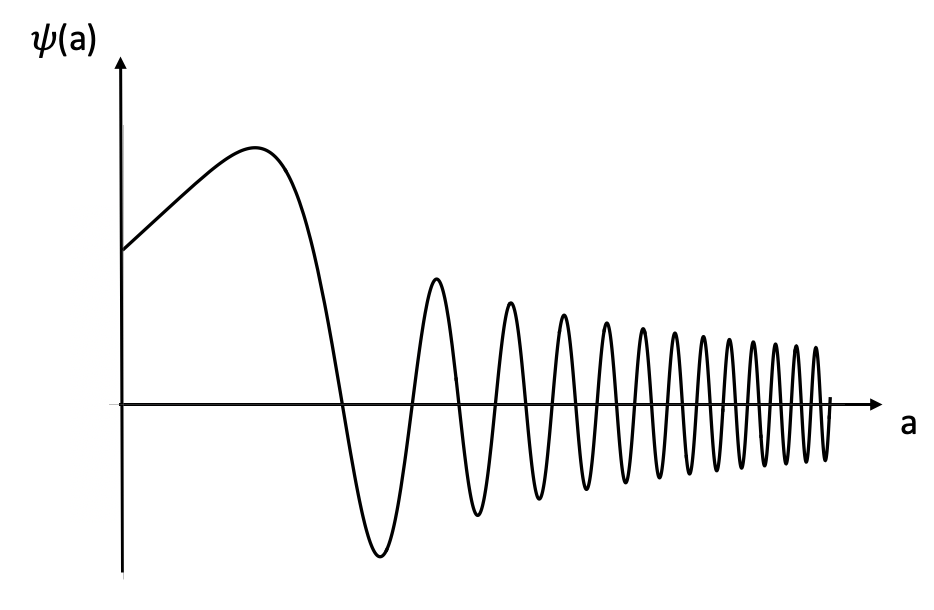}
\\ Figure 1: \emph{Wave Function of the Universe as a function of the scale factor}
\end{center}
\subsubsection{The case $\alpha = 0$}
Setting $\alpha = 0$ in  Lagrangian \eqref{specificlagra}, the Hamiltonian becomes
\begin{equation}
\mathcal{H} = -6 \beta  k a -\frac{\pi_a^2}{24 \beta  a}
\end{equation}
and the related WDW equation can be solved even without imposing $k=0$. It  provides
\begin{equation}
\label{wfbeta}
    \psi(a,Q) =  \psi_0 \, \left[c_1 D_{-\frac{1}{2}}\left(2 i \sqrt{6\beta \sqrt{-k}} \, a \right) + c_2 D_{-\frac{1}{2}}\left(2  \sqrt{6\beta \sqrt{-k}} \, a \right)\right],
\end{equation}
with $i$ being the imaginary unit and $D$ are parabolic cylinder functions. The wave function becomes trivial in the case of a spatially-flat universe with $k=0$. From Eq. \eqref{wfbeta},
by applying the Hartle Criterion and pursuing the same approach as in the previous subsection, we  get  classical solutions. In particular, as showed in  Fig. 2, we notice that the parabolic cylinder functions asymptotically behave like the exponential function; therefore, taking the real part and considering that the wave function consists of a linear combinations of functions with subscript $-1/2$, the Hamilton-Jacobi action, up to a total factor, can be written as:
\begin{equation}
    S_0 \sim 6 \beta \sqrt{-k} a^2,
\end{equation}
with a conjugate momentum given by
\begin{equation}
    \pi_a = \frac{\partial S_0}{\partial a} \sim 12 \beta \sqrt{-k} a.
    \label{momentums0}
\end{equation}
By equating Eq. \eqref{momentums0} with 
\begin{equation}
    \pi_a = \frac{\partial \Lagr}{\partial \dot{a}} = - 12 \beta a \dot{a}\,,
\end{equation}
we get the differential equations
\begin{equation}
    \sqrt{-k} - \dot{a} = 0,
\end{equation}
whose solution is
\begin{equation}
a(t)=  \sqrt{-k} t + c,
\end{equation}
in  agreement with results reported in Eq. \eqref{alpha0solutions}. Once again, it is worth stressing that the Hartle Criterion can be used to find classical solutions without directly solving the field equations. 
\begin{center}
\centering
\includegraphics[width=.55\textwidth]{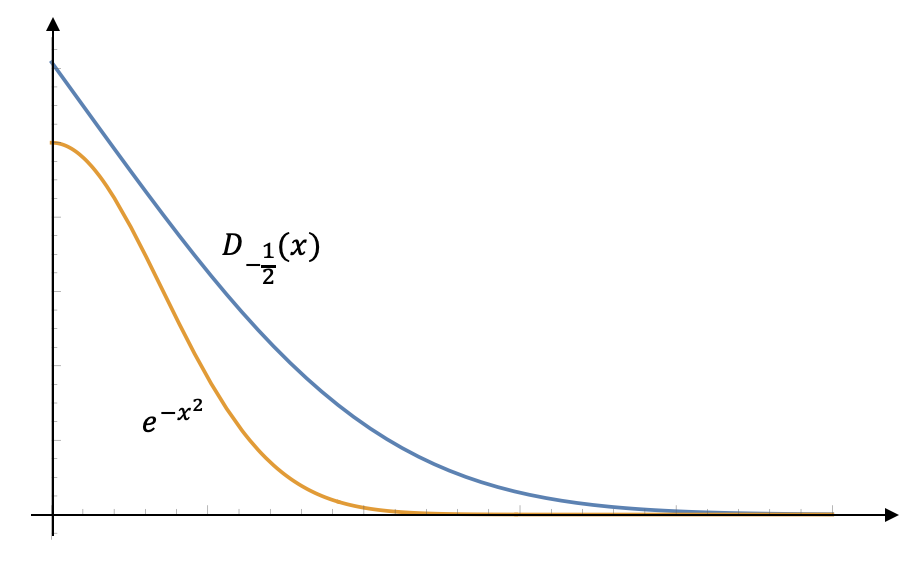}
\\ Figure 2: \emph{Comparison between exponential function (orange line) and parabolic cylinder function (blue line)}
\end{center}
\subsubsection{The case $\beta = 0$}
Another interesting case, which deserves to be studied separately, is the case $\beta = 0$. Here, the only analytic solution occurs for $k=0$, where the Hamiltonian becomes
\begin{equation}
 \mathcal{H} = - \frac{Q^{1-s}}{24 \alpha  s \, a} \pi_a^2 +\alpha 
   (s-1) a^3 Q^s
\end{equation}
and, therefore, the wave function turns out to be
\begin{equation}
     \psi(a,Q) =  \psi_0 \, \sqrt{a} \, \left[\alpha^2  s (1-s) Q^{2s-1}\right]^{1/12} \left[c_1 J_{\frac{1}{6}}\left(a^3 \alpha\sqrt{ s (1-s) Q^{2s-1}}\right) +c_2 J_{-\frac{1}{6}}\left(a^3 \alpha \sqrt{s (1-s) Q^{2s-1}}\right)\right] .
\end{equation}
This case is  interesting because it is possible to suitably change the variables  to introduce a cyclic coordinate into the Minisuperspace derived from the Noether symmetries \cite{Book}. Specifically, we can operate the coordinate transformation
\begin{equation}
\{a,Q\}\longrightarrow \{z,w\}\,,
\end{equation}
which explicitly is
\begin{equation}
 a(t) = \left(\frac{3z(t)}{2s}\right)^{2 s/3},\qquad Q(t) = \frac{4 s^2 \omega(t)^{\frac{1}{s}}}{9 z(t)^2},
 \label{inversion}
\end{equation}
the Lagrangian becomes
\begin{equation}
\Lagr = -\alpha  \omega \left(6 s \, \omega^{-1/s} \dot{z}^2+s-1\right),
\end{equation}
which is cyclic with respect to $z$. After a  Legendre transformation,  we obtain
\begin{equation}
{\cal H} = \omega(s-1) + \frac{3 \pi^2_z}{24 s \omega^{\frac{s-1}{s}}}\,,
\end{equation}
where $\pi_z$ is the conjugate momentum of the new cyclic variable $z(t)$. Now, since $z$ is cyclic, we  have
\begin{equation}
    \frac{d\pi_z}{dt} = 0\,, 
\end{equation}
meaning that $\pi_z$ is a constant:
\begin{equation}
    \pi_z \equiv \Sigma_0\,.
\end{equation}
Therefore, the Wave Function of the Universe is  a  solution of the  following differential equations:
\begin{equation}
\begin{cases}
\displaystyle {\cal H} \psi = 0
\\
\displaystyle i\partial_z \psi = \Sigma_0 \psi, 
\end{cases}
\end{equation}
where  the first is the  WDW equation and the second is related to the momentum conservation. The general solution  is
\begin{equation}
\psi(z,\omega) = \psi_0 \exp \left\{i\left[2 \omega^{\frac{2s -1}{2s}} \sqrt{2s(s-1)}\right]z\right\} \;,
\end{equation}
which is an oscillating wave function satisfying the Hartle Criterion as follows. Specifically, adopting the semiclassical limit \cite{Maniccia:2022iqa} and recasting $\psi$ in terms of the action $S_0$ as
\begin{equation}
\psi \sim \exp\left\{i S_0 \right\} \;,
\end{equation}
the Hamilton-Jacobi equations can provide back the equations of motion for the new variables. More precisely, the system of Hamilton-Jacobi equations
\begin{equation}
\begin{cases}
\displaystyle \frac{\partial S_0}{\partial z} = \pi_z = \Sigma_0
\\
\\
\displaystyle \frac{\partial S_0}{\partial w} = \pi_w = 0 \;,
\end{cases}
\label{ELWF}
\end{equation} 
yields
\begin{eqnarray}
&& s w \ddot{z} + \dot{w} \dot{z} (s-1) = 0 ,
\\
&& w = - 6^s  \dot{z}^{2s}\,.
\label{ELw}
\end{eqnarray} 
It can be easily shown that the system \eqref{ELWF} is equivalent to the system of Euler-Lagrange equations with respect to $a$ and $Q$, when coming back to the old variables. As a matter of facts, by considering the inverse transformations of Eq. \eqref{inversion}, that is
\begin{equation}
    w = a^3 Q^s , \qquad z = \frac{2s}{3} a^{\frac{3}{2s}},
\end{equation}
the second equation of \eqref{ELw} becomes
\begin{equation}
    a^3 Q^s = -6^s a^{3-2s}\dot{a}^{2s},
\end{equation}
from which
\begin{equation}
    Q = - 6 \frac{\dot{a}^2}{a^2},
\end{equation}
namely the Euler--Lagrange equation with respect to $Q$. Equivalently, after few manipulations, the first equation of \eqref{ELw} takes the form:
\begin{equation}
    (2 s-1) a \ddot{a} +(3-2 s) \dot{a}^2 = 0
\end{equation}
which is the equation of motion with respect to the scale factor.
Notice finally that, also here, the Wave Function of the Universe is oscillating and therefore the Hartle Criterion is satisfied. 

\section{ Conclusions}
\label{concl}
In this work we considered the modified STEGR model $f(Q) = \beta Q + \alpha Q^s$ and investigated  the related Minisuperspace Quantum Cosmology after discussing  the FLRW cosmology  and  the related  energy conditions. The WEC is identically violated and the further degrees of freedom of $f(Q)$ gravity can be dealt as a geometric perfect fluid behaving as dark energy.

The starting point for  Quantum Cosmology  is the Hamiltonian formulation. 
From this, it is possible to achieve the WDW equation and then the Wave Function of the Universe.
It  can provide information on the early stages of the universe  after the application of the Hartle Criterion. Interestingly, we found that the wave function trend is oscillating only for large scale factors, that is in the classical zone  related to observable universes, while it keeps an exponential trend at early times. This confirms the validity of the Hartle Criterion for this model. In this regard, we showed that a suitable change of variables allows to recast the Wave Function of the Universe as $\psi \to e^{i S_0}$, with $S_0$ being a classical action. Therefore, by means of the Hamilton-Jacobi equations, the cosmological solutions can be straightforwardly recovered even without solving the field equations directly. Clearly, due to the absence of a self-consistent theory of Quantum Gravity, this approach represents a first step and a sort of effective model, which can be useful for cosmological purposes at  early times. Moreover, modifications of STEGR can be extremely helpful to fix  shortcomings suffered by GR, since inflationary behaviors and dark energy effects  arise  without introducing any cosmological constant, but  provided by additional geometric contributions in the field equations. This prescription can be also applied to the $f(R)$ extension of GR but, unlike $f(R)$ gravity, modified STEGR leads to second-order field equations, which can be handled more easily. From this point of view, $f(Q)$ gravity can be useful to fix several problems of GR at different scales of energy.

\section*{Acknowledgments}

The Authors acknowledge the support of {\it Istituto Nazionale di Fisica Nucleare} (INFN) ({\it iniziativa specifica}   QGSKY). 
This paper is based upon work from the COST Action CA21136, \textit{Addressing observational tensions in cosmology with systematics and fundamental physics} (CosmoVerse) supported by COST (European Cooperation in Science and Technology).

\appendix
\section{The Arnowitt-Deser-Misner Formalism and the Wheeler-De Witt Equation}
\label{ADMApp}
Let us briefly introduce the Arnowitt-Deser-Misner (ADM) formalism for the Hamiltonian formulation of GR and  the WDW equation.  After the formulation of Einstein's theory gravity, several attempts have been pursued to the purpose of developing a quantum theory of gravity where  Quantum Mechanics and GR could be accounted under the same standard \cite{DeWitt:1967yk, DeWitt:1967ub}. One of the first attempt in this direction, is represented by the ADM formalism, developed  in 1962 \cite{Arnowitt:1962hi}. This formalism aims to construct a quantization scheme for GR, with a corresponding wave function capable of addressing the evolution of  metric degrees of freedom in a quantum picture. In particular, denoting the three-dimensional space variables with Latin indexes, the Hilbert-Einstein action can be written  as~\cite{DeWitt:1967yk, Thiemann:2001gmi}:
\begin{equation}
S = \frac{1}{2} \int_V \sqrt{-g} \, R \, d^4x + \int_{\partial V} \sqrt{h} K \; dx^3\,,
\label{ADMAction}
\end{equation}
where $V$ is the four-dimensional manifold, $\partial V$ the three-dimensional boundary and $K$ is the trace of the extrinsic curvature tensor $K_{ij}$, namely $ K = h^{ij} K_{ij}$, with $h_{ij}$ being the three-dimensional spatial metric.

Therefore, the metric can be decomposed such that the evolution of its spatial part can be thought as the evolution of hypersurfaces along the time coordinate according to the so called \textit{Geometrodynamics}
\cite{Hartle:1983ai}. According to this interpretation, the Hilbert-Einstein Lagrangian density $\mathscr{L}$ can be recast in terms of the lapse function $N$, the extrinsic curvature $K_{ij}$ and the extrinsic curvature $^{(3)}R$ as:
\begin{equation}
\mathscr{L}=\frac{1}{2} \sqrt{h} N\left(K^{i j} K_{i j}-K^{2}+{ }^{(3)} R\right) + t. d. \,,
\label{lagradens}
\end{equation}
Starting from Eq.\eqref{lagradens},  one can get the Poisson brackets for the Hamiltonian and the three-dimensional metric $h_{ij}$. Adopting a  canonical quantization scheme, the Poisson brackets can be promoted to commutators, providing the relations:
\begin{equation}
\begin{cases}
[\hat{h}_{ij}(x), \hat{\pi}^{kl} (x')] = i \; \delta^{kl}_{ij} \; \delta^3 (x-x')\,,
\\
\delta^{kl}_{ij} = \frac{1}{2} (\delta_i^k \delta_j^l + \delta_i^l \delta_j^k)\,,
\\
[\hat{h}_{ij}, \hat{h}_{kl}] = 0\,,
\\
[\hat{\pi}^{ij}, \hat{\pi}^{kl} ] = 0\,.
\end{cases}
\end{equation}
where $\pi^{ij}$ is the momentum conjugated to the metric $h_{ij}$, defined as
\begin{eqnarray}
\pi^{i j} \equiv \frac{\delta \mathscr{L}}{\delta \dot{h}_{i j}}=\frac{\sqrt{h}}{2}\left(K h^{i j}-K^{i j}\right).
\label{conjmom}
\end{eqnarray}
The condition of null energy for gravitational field, $\mathcal{H} = 0$, becomes the quantum equation
\begin{equation}
\hat{{\cal{H}}}| \psi> = 0\,,
\end{equation}
where $\hat{{\cal{H}}}$ is the quantum Hamiltonian which can be expressed in terms of the \textit{Superlaplacian} operator $\nabla^2$ and the extrinsic curvature as: 
\begin{equation}
{\cal{H}} = \left(\nabla^2 - \frac{\kappa^2}{4} \sqrt{h} \; ^{(3)} R \right)\,,
\label{constraints}
\end{equation}
with
\begin{equation}
\nabla^2 = \frac{1}{\sqrt{h}}\left(h_{i k} h_{j l}+h_{i l} h_{j k}-h_{i j} h_{k l}\right) \frac{\delta}{\delta h_{ij}}\frac{\delta}{\delta h_{kl}}\,.
\end{equation}
Here $\psi$ is the \textit{Wave Function of the Universe}  describing the evolution of the gravitational field, defined in the infinite-dimensional  \textit{Superspace} of configurations of all possible three-metrics. Due to the impossibility of dealing with the infinite-dimensional Superspace, the WDW equation cannot be generally solved. Moreover, unlike Quantum Mechanics, in the ADM formalism the quantity $|\psi|^2$ is not definite positive, with the consequence that a self-consistent definition of Hilbert space cannot be achieved. In addition, the probabilistic interpretation of the wave function as a probability amplitude does not apply here.

Nevertheless, the standard procedure to address these issues consists of reducing the Superspace to a \textit{Minisuperspace}, a finite-dimensional configuration space where the WDW equation can be solved and the wave function  provides information. In other words, instead of solving the infinite-dimensional functional WDW equation, it is possible to reduce the problem to a finite-dimensional one, quantize and solve the related dynamical system. Specifically, in cosmology, such Minisuperspace are the configuration space of the scale factor and scalar fields sourcing the cosmological equations.  

 While waiting for a self-consistent  theory of Quantum Gravity, Minisuperspace Quantum Cosmology  represents an effective approach capable of providing information for early universe. In this context, the meaning of the wave function plays a key role, in particular in the so-called  \emph{Many World Interpretation} of Quantum Mechanics stating that the wave function  comes from quantum measurements  simultaneously realized in different universes, without exhibiting any collapse as in standard Quantum Mechanics~\cite{Bousso:2011up}. Another possibility is that the wave function can be related to the probability for the early universe to evolve towards  classical universes like the one we are living in ~\cite{Vilenkin:1988yd, Hartle:1983ai, Hawking:1983hj}. 

\section{The Hartle Criterion as a selection rule for observable universes}
\label{HartleSec}

An interesting attempt to interpret the meaning of  the Wave Function of the Universe has been provided by J.~B.~Hartle, who proposed a criterion based on the analogy with Quantum Mechanics. In particular, according to Hartle, the wave function  keeps an oscillating behaviour in the classically allowed region of configuration space, \emph{i.e.} where classical universes occur because cosmological parameters are correlated, and an exponential behaviour where correlations are not present ~\cite{Hartle:1983ai}. In the second case, the wave function is an \textit{instanton} solution with Euclidean time.

This scheme of interpretation is inspired to standard non-relativistic Quantum Mechanics, where the wave function of free particles is oscillating outside the barrier and exponential inside.

On the other hand, it is possible to show that such a criterion is a \textit{selection rule} if Noether symmetries exist for the related Minisuperspace \cite{Gaetano}. In this case, the oscillation of the Wave Function of the Universe and then the emergence of classical observable universes is related to the existence of conserved quantities.

Considering the above interpretations and recasting the Wave Function of the Universe as $\psi \sim e^{i S_0}$, classical solutions to the field equations can be straightforwardly recovered from the Hamilton-Jacobi equations. Therefore,  classical solutions can be obtained even without solving the field equations directly. This is particularly important in modified theories of gravity where  dynamics  cannot often be analytically solved. From this point of view, the Hartle criterion can be adopted to recover classical trajectories in the configuration space, namely solutions to the field equations in terms of the scale factor, denoting the classical evolution of the universe.

In order to show the strict relation occurring between Hartle criterion and conserved quantities, we start from the semiclassical expression of the wave function, that is $\psi = e^{i m_P^2 S}$, where $m_P$ accounts for the Planck mass and $S$ is the classical action. 

If the system shows  symmetries, then it is possible to introduce cyclic variables and conserved momenta. Therefore, if $\pi^i_{q^i}$ are the momenta related to the Minisuperspace variables $q^i$ and $j_i$ the conserved quantities,  we have
\begin{equation}
j_i \equiv \pi^i_{q^i} = \frac{\partial \Lagr}{\partial \dot{q}^i},
\label{jcosmo}
\end{equation}
with $q^i$ being cyclic. Once the conjugate momenta are promoted to quantum operator, the above relation becomes
\begin{equation}
- i \partial_i |\psi> = j_i |\psi>.
\label{jcosmoquant}
\end{equation}
If the system carries $m$ symmetries (and corresponding conserved quantities $j_1, j_2 \dots j_m$), Eq. \eqref{jcosmoquant} and the WDW equation provide a system of $m+1$ differential equations:
\begin{equation}
\begin{cases}
\hat{\mathscr{H}} |\psi>=0
\\
- i \partial_1 |\psi> = j_1 |\psi>
\\
- i \partial_2 |\psi> = j_2 |\psi>
\\
\mathrel{\vdots}
\\
- i \partial_m |\psi> = j_m |\psi>.
\end{cases}
\label{systemconstraint1}
\end{equation}
Since $j_i$ are conserved quantities, the wave function will be oscillating in the Minisuperspace of variables $q^i$; in particular, the latter $m$ equations of the above system provide the following solution \cite{Gaetano, Capozziello:2012hm}:
\begin{equation}
|\psi> = e^{i j_k q^k} |\chi(q^\ell)>, \qquad m < \ell < n
\end{equation}
where $m$ is the number of symmetries, $\ell$ are the directions where symmetries do not exist, $n$ is the total dimension of Minisuperspace. Notice that also the component $|\chi>$ can generally depend on the conserved quantities $j_i$, though its functional behavior depends on the specific case considered. In summary, the Hartle Criterion can be  related to symmetries, which, in turn, can select  classical solutions which can be interpreted as observable universes.

\end{document}